# Sensitivity study of (10,100) GeV gamma-ray bursts with double shower front events from ARGO-YBJ *

Xun-Xiu Zhou (周勋秀)[1; 1)]  Lan-Lan Gao (高兰兰)[1]  Yu Zhang (张宇)[1]  Yi-Qing Guo (郭义庆)[2]
Qing-Qi Zhu (朱清棋)[2]  Huan-Yu Jia (贾焕玉)[1]  Dai-Hui Huang (黄代绘)[1]

[1] School of Physical Science and Technology, Southwest Jiaotong University, Chengdu 610031, China
[2] Institute of High Energy Physics, CAS, Beijing 100049, China

**Abstract:** ARGO-YBJ, located at the YangBaJing Cosmic Ray Observatory (4300 m a.s.l., Tibet, China), is a full coverage air shower array, with an energy threshold of ~300 GeV for gamma-ray astronomy. Most of the recorded events are single front showers, satisfying the trigger requirement of at least 20 particles detected in a given time window. However, in ~11.5% of the events, two randomly arriving showers may be recorded in the same time window, and the second one, generally smaller, does not need to satisfy the trigger condition. These events are called double shower front events. By using these small showers, well under the trigger threshold, the detector primary energy threshold can be lowered to a few tens of GeV. In this paper, the angular resolution that can be achieved with these events is evaluated by a full Monte Carlo simulation. The ARGO-YBJ sensitivity in detecting gamma-ray bursts (GRBs) by using double shower front events is also studied for various cutoff energies, time durations, and zenith angles of GRBs in ARGO's field of view.

**Keywords:** double shower front events, angular resolution, GRBs, sensitivity, ARGO-YBJ

**PACS:** 98.70. Rz

## 1. Introduction

Gamma-ray bursts (GRBs), an energetic form of energy released from unpredictable cosmic locations, have been one of the most captivating astronomical phenomena since their first discovery. Although thousands of GRBs detected by satellite-based experiments such as Swift [1], HETE [2] and Fermi-GBM [3] are concentrated in the keV−MeV energy range, EGRET [4] and Fermi-LAT [5] have observed photons in the MeV−GeV range. Up to the time of writing (October 2015), about 60 GRBs have been detected to contain photons with energy above 1 GeV, of which 14 GRBs have their highest observed photon energy above 10 GeV [6, 7]. Particularly, Fermi-LAT announced the detection of the highest recorded photon energy (95 GeV) from GRB 130427A [8]. So far the redshifts of more than 400 GRBs have been observed. The lowest recorded redshift of GRBs is lower than 0.01[9] and the highest

* Supported by National Natural Science Foundation of China (Grant No. 11475141) and the Fundamental Research Funds for the Central Universities (Grant No. 2682014CX091)
1) E-mail: zhouxx@swjtu.edu.cn

GRBs is 8.1 from GRB 090423 [10]. Redshift measurements show that these GRBs occurred at cosmological distances.

Though much progress has been achieved from satellite-based experiments, lots of basic questions, such as the emission mechanisms of GRBs, still remain unresolved [8]. Due to the limited size of the detectors on board satellites and the rapid fall of GRB energy spectra, satellite-based experiments hardly cover the energy region above 1 GeV. In order to see the complete picture, it is important to improve the sample statistics of high energy GRBs. Ground-based experiments including extensive air shower (EAS) arrays and imaging atmospheric Cherenkov telescopes (IACTs) can easily reach much larger effective areas, and thus are especially suitable for $E > 10$ GeV GRB detection. Searches for high energy emission above 10 GeV from GRBs have been done by ground-based experiments such as MAGIC [11], VERITAS [12], HESS [13], ARGO-YBJ [14, 15] and Tibet AS$\gamma$ [16]. No significant detection of high energy emission from GRBs has been observed so far, although some positive indications have been reported. Now, several ground-based experiments are attempting to improve their sensitivity in detecting GRBs by reducing the energy threshold [17, 18].

The single particle technique (SPT) allows the energy threshold of ARGO-YBJ to be lowered to ~1 GeV [19], but this technique does not provide information about the energy and arrival direction of the primary gamma rays. By using double shower front events, the primary energy threshold of the ARGO-YBJ detector can be lowered to a few tens of GeV. In this work, we study the angular resolution of double shower front events and the sensitivity of the ARGO-YBJ detector for GRBs detection by using these events. This paper is organized as follows. The ARGO-YBJ experiment and double shower front events are introduced in Section 2. The angular resolution of double shower front events is shown in Section 3. The sensitivity for GRBs is presented in Section 4. The conclusion is given in Section 5.

## 2. ARGO-YBJ experiment and double shower front events

The ARGO-YBJ experiment, a collaboration among Chinese and Italian institutions, is an EAS detector located in Tibet, China at an altitude of 4300 m above sea level. It is mainly devoted to VHE gamma-ray astronomy and cosmic ray physics. The detector is composed of a single layer of resistive plate chambers (RPCs), operated in streamer mode, with a modular configuration. The basic module is a cluster (5.7×7.6 m$^2$), composed of 12 RPCs. Each RPC is read out by 10 pads, representing the space-time pixels of the detector. The clusters are disposed in a central full-coverage carpet (130 clusters on an area 74×78 m$^2$). In order to improve the performance of the apparatus in determining the shower core position, the central carpet is surrounded by 23 additional clusters ("guard ring"). More details of the detector can be found elsewhere [20].

The ARGO-YBJ detector is connected to two independent data acquisition systems, corresponding to two operation modes [19]. One is the scaler mode which counts the single particle rate. The other is the shower mode where ARGO-YBJ detector is operated by requiring at least 20 particles within 420 ns on the entire carpet detector. For each triggered event ("SF Event"), all the hits registered within a time window of 2134 ns are recorded. During this time, another randomly arriving shower (of which pad-multiplicity is more than 5）, with a probability of about 11.5% (from ARGO-YBJ data), may be recorded. These coincident events are called double shower front events ("DF Events") in this paper. Figures 1(a) and (b) show the arrival time of particles measured by time to digital converters (TDCs) for a normal triggered event and a double front shower event. Figure 2 shows the space-time pixels (TDC values vs. pad coordinates) of secondary particles. A larger TDC value means an earlier registered hit. The TDC times of a triggered event are distributed from 1150 ns to 1450 ns. The double front shower event, recorded in other times within the 2134 ns, is normally the smaller shower event. The double front shower event is selected with the following steps: 1) according to the TDC time, find the triggered event which arrives first and satisfies the trigger requirement of at least 20 particles; 2) Beyond about 200 ns from the triggered event, accept it as a double front shower event if there are more than 5 pads hit within 100 ns.

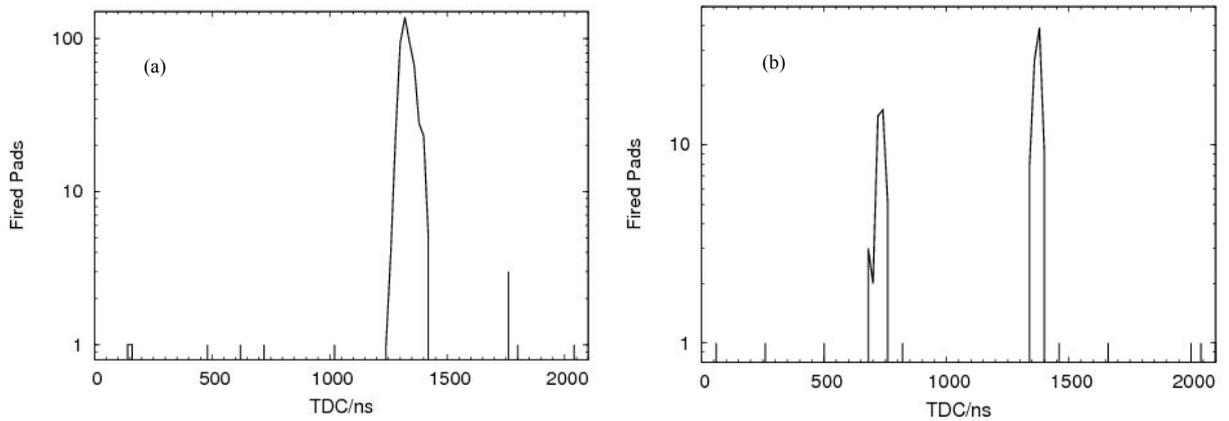

Fig. 1. The distribution of TDC values of secondary particles in ARGO-YBJ arrays for a normal triggered event (a) and a double shower front event (b).

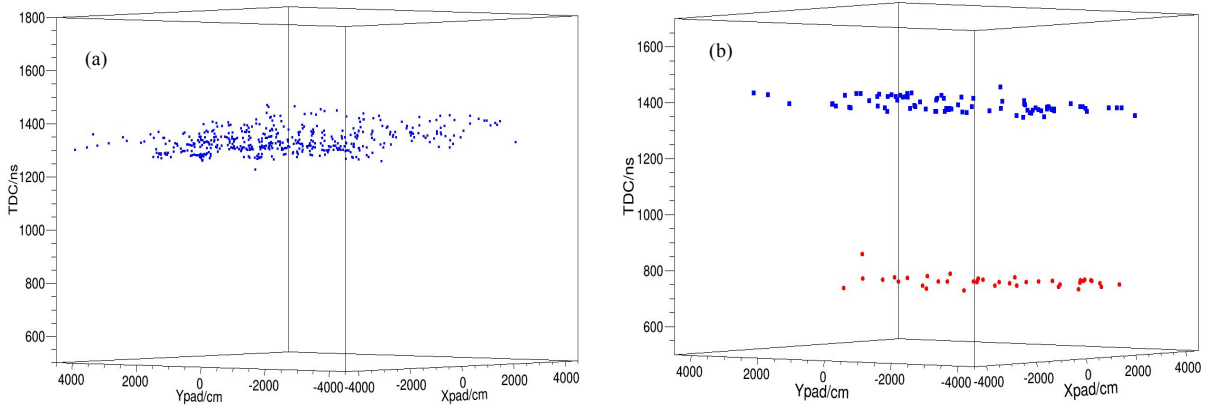

Fig. 2. TDC values vs. pad coordinates of secondary particles for a normal triggered event (a) and a double shower front event (b).

Figure 3 shows the event rate distribution with ARGO-YBJ data taken from 28[th] March 2011. The event rate is ~3330 Hz for triggered events and ~382 Hz for double shower front events.

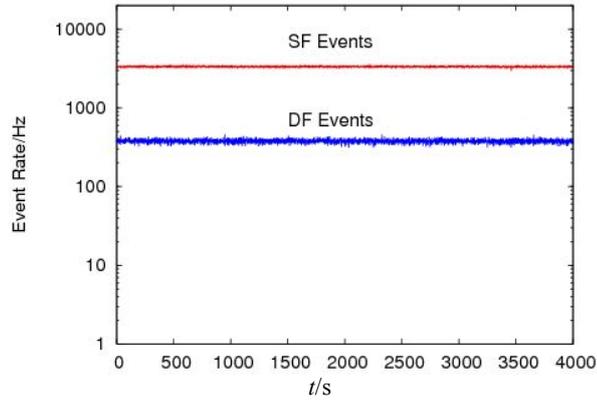

Fig. 3. Event rate distribution for triggered events and double shower front events.

Figure 4 shows the pad multiplicity ($N_{pad}$) distribution of triggered events and double shower front events. Since there is no need for the smaller shower to satisfy the trigger requirement, the energy threshold can be lowered by using double shower front events.

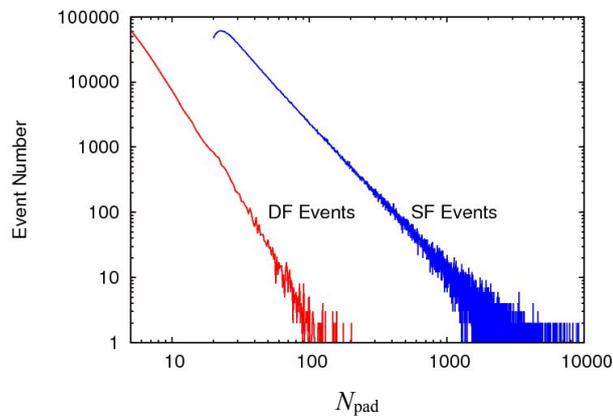

Fig. 4. $N_{pad}$ distribution for triggered events and double shower front events.

## 3. Simulation of double shower front events and angular resolution

In this work, we use CORSIKA [21] to simulate the evolution and properties of extensive air showers in the atmosphere. Because double shower front events cannot be simulated by using the normal G4argo [22], the data which is generated by CORSIKA7.3700 should be mixed according to the following principles:

1) Estimate the average number of events ($\lambda$) within the time window (2134 ns) through the fluence of cosmic rays.

2) Sample the number of events in 2134 ns subject to a Poisson distribution based on the parameter $\lambda$, with the time interval between two events sampled on the basis of an exponential distribution.

3) Rank the time of secondary particles produced by CORSIKA in accordance with the mid arriving time of intensive events.

4) Put another photon at random into the time window in the simulation of photons.

Next, the mixed data is input to G4argo, which is based on GEANT4 [23] and used to simulate the response of the ARGO-YBJ detector. Here, the form of input and output in the normal G4argo is modified to be available for double shower front events. Lastly, double shower front events are selected using the procedure presented in the last section, and then reconstructed using the least squares method with conical fit [24]. The sampling area is 200 m×200 m around the carpet center.

Figure 5 shows the comparison of zenith angle distribution of double shower front events between Monte Carlo simulation samples and ARGO-YBJ data. Due to atmospheric absorption, the efficiency is much lower for higher zenith angle $\theta$ [14]. We can see the distribution of simulation data is consistent with experimental data within the 5% level at $\theta < 45°$.

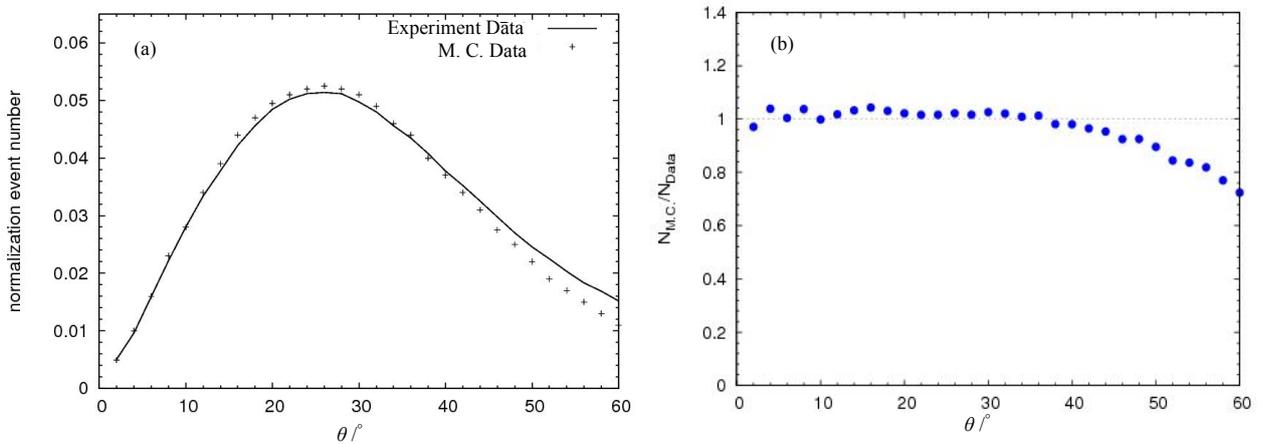

Fig. 5. Zenith angle distribution of double shower front events between M. C. data and ARGO-YBJ data.
(a): solid line for ARGO-YBJ data and cross points for M.C. samples. (b): circle points for the ratio of M.C. sample to ARGO-YBJ data.

$\Psi_{72}$ (the angular resolution) is the angular aperture with respect to the simulated shower direction containing 72% of the reconstructed tracks. In Fig. 6, we show the angular resolution of double shower front events as a function of pad multiplicity for both photons and protons. The zenith angle is chosen to vary from 0° to 60°. The value of the angular resolution for photon-induced showers is slightly better than that for proton-induced showers.

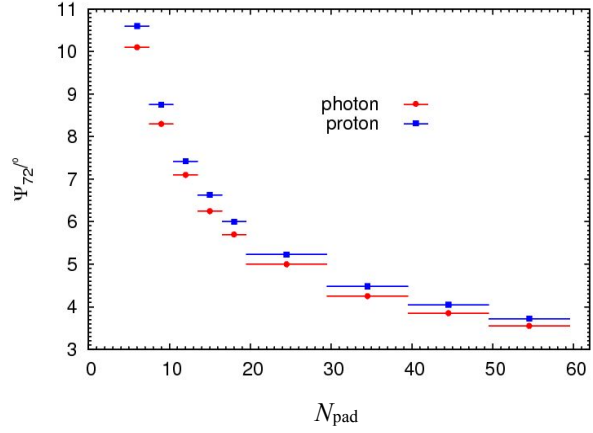

Fig. 6. Angular resolution of double shower front events as a function of $N_{pad}$ for both photons and protons (the horizontal bars refer to the width of the $N_{pad}$ bins).

The angular resolution of double shower front events is a little worse than that of triggered events [15], one reason for which is that some hits in triggered events are blended in double shower front events.

## 4. ARGO-YBJ sensitivity to GRBs using double shower front events

### 4.1 Effective area of ARGO-YBJ apparatus in observing GRBs

The ARGO-YBJ sensitivity in detecting a GRB is determined by the effective area and the angular resolution. The effective area, $A_{eff}$, is calculated by means of Monte Carlo simulation, which is identical to the simulation of the angular resolution. $A_{eff}$ depends on the primary energy $E$ and the zenith angle $\theta$. We calculate $A_{eff}$ at $E = 10^1, 10^{1.2}, 10^{1.4}, 10^{1.6}, 10^{1.8}$ and $10^2$ GeV, for $\theta = 0°, 10°, 20°$ and $30°$. By definition, $A_{eff}$ can be expressed as:

$$A_{eff}(E,\theta) = \frac{n_s}{N} \cdot A_s \cdot \cos\theta . \qquad (1)$$

Here, $n_s$ is the number of successfully reconstructed double shower front events, $N$ is the total number of events generated by CORSIKA, and $A_s$ is the sampling area ($200\text{m} \times 200\text{m}$). Figure 7 shows the $A_{eff}$ of photons for double shower front events as a function of primary energy, where each line corresponds to a fixed zenith angle. The effective area increases with the decrease of zenith angle and the increase of primary energy. At $\theta = 10°$, the effective area is about 0.025 m² at $E$=10 GeV and about 5.684 m² at $E$=100 GeV. The $A_{eff}$ of photons for double shower front events is consistent with the results in [25].

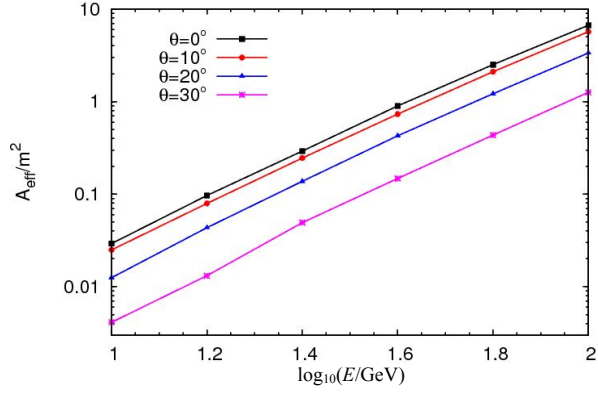

Fig. 7. The $A_{eff}$ of photons for double shower front events at different zenith angles as a function of the primary energy.

## 4.2 Sensitivity of ARGO-YBJ in searches for GRBs

In the ARGO-YBJ experiment, a GRB appears as a shower cluster in a given small sky window and a time interval with an appropriate significance [26]. In this paper we take 5σ as the necessary significance to specify a new discovery of a GRB from the background fluctuation. From the above Monte Carlo simulation of double shower front events we know that the angular resolution ($\Psi_{72}$) is about 7.9° for photons with E ≥10 GeV. In this work, the GRB search is performed in an angular window with radius 7.9°. The "equi-zenith-angle" method [27] is used to estimate the background event rate from ARGO-YBJ data. Within a time duration of 1 second, the number of background events $<N_b>$ is the average number from 10 off-source windows (the radius of each angular window chosen as 7.9°) in the azimuthal directions with the same zenith angle. $N_{on}$, which is the number of events falling within the on-source window (same size as off-source window) in a given time interval, can be calculated under a 5$\sigma$ observation by following Eq. (2) and Eq. (3).

The probability ($P_b$) of a candidate event being induced by background fluctuation can be calculated by [27]:

$$P_b = \sum_{i=N_{on}+1}^{\infty} P(i) + \frac{1}{2} P(N_{on}). \qquad (2)$$

Here, $P(i)$ is the Poisson probability for the multiplicity $i$ and a given $<N_b>$. A small value of $P_b$ means a high possibility of GRBs. $P_b$ can be transformed into the significance of a Gaussian distribution ($S$) as follows:

$$P_b = \int_S^{\infty} \frac{1}{\sqrt{2\pi}} \cdot e^{-\frac{1}{2}x^2} dx. \qquad (3)$$

Considering the angular resolution, the minimum number of signal events $N_s$ within 1 second can be calculated by: $N_{on} = N_s \times 72\% + \langle N_b \rangle$. Figure 8 shows the values of $\langle N_b \rangle$ and $N_s$ for double shower front events.

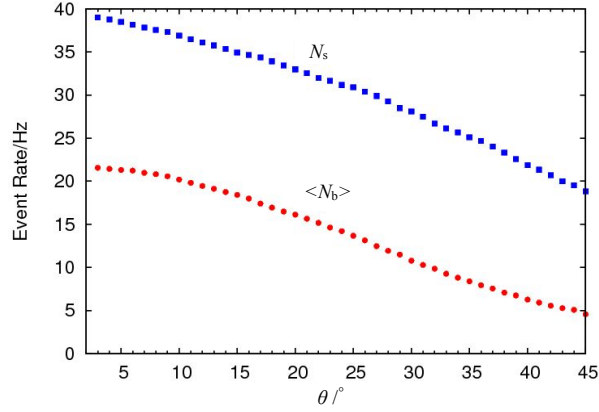

Fig. 8. The background event rate $\langle N_b \rangle$ and the minimum signal event rate $N_s$ as a function of $\theta$ for double shower front events (a GRB signal with a rate equal to or higher than $N_s$ is observed with a significance at least of $5\sigma$.).

The normalization constant $K$ can be calculated by

$$N_s(\theta, \Delta t) = K \int_{E_{min}}^{E_{max}} A_{eff}(E, \theta) \cdot E^{-\alpha} \cdot e^{-\tau(E,z)} dE. \qquad (4)$$

Here, $N_s$ is the minimum number of signals, $E_{max}$ is the energy cutoff, $A_{eff}$ is from Eq. (1), and $\tau(E, z)$ is the optical depth due to the extragalactic background light (EBL) absorption. We use the value of $\tau(E, z)$ from the Gilmore Model (2012) [28, 29], and fix $E_{min}$ at 10 GeV in this paper. The power law index $\alpha$ of primary gammas is assumed to be -2.0.

The fluence $F$ (from 10 GeV to $E_{max}$) required for a $5\sigma$ observation can be calculated as follows:

$$F = K \int_{10\,\text{GeV}}^{E_{max}} E \cdot E^{-\alpha} \cdot dE. \qquad (5)$$

Here, $K$ is from Eq. (4). Assuming $\alpha = -2.0$, $\theta = 10°$, and $\Delta t = 0.1\text{s}, 1\text{s}, 10\text{s}$, respectively, nine curves in Fig. 9 show the $5\sigma$ minimum fluence as a function of $E_{max}$ for $z = 0.0, 0.5, 1.0$.

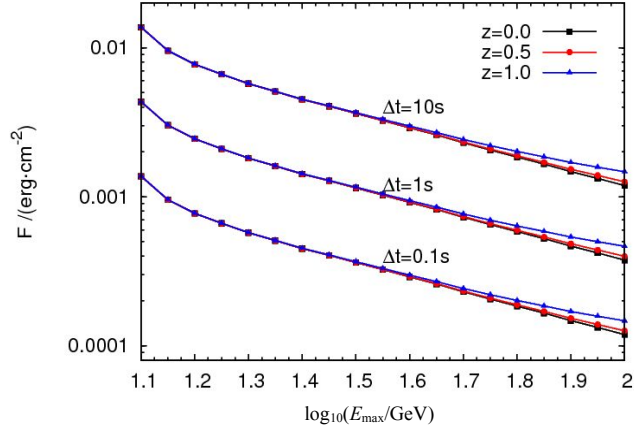

Fig. 9. The discovery fluence $F$ (for double shower front events) as a function of $E_{max}$ for different redshifts and time durations.

It can be seen from Fig. 9 that the ARGO-YBJ sensitivity is better in the case of a high energy cutoff and short time duration. With the energy increasing, the EBL effect for gamma-rays strengthens. Therefore, the decrease in the energy threshold enhances the sensitivity in detecting GRBs.

Figure 10 shows the discovery fluence $F$ for double shower front events as a function of $E_{max}$ at $z = 0.0$ and $\Delta t = 1$s, where each line corresponds to a fixed zenith angle. According to Fig. 10, the $5\sigma$ minimum fluence of double shower front events, which characterizes the sensitivity of detecting GRBs, varies from $10^{-4}$ erg.cm$^{-2}$ to $10^{-2}$ erg.cm$^{-2}$.

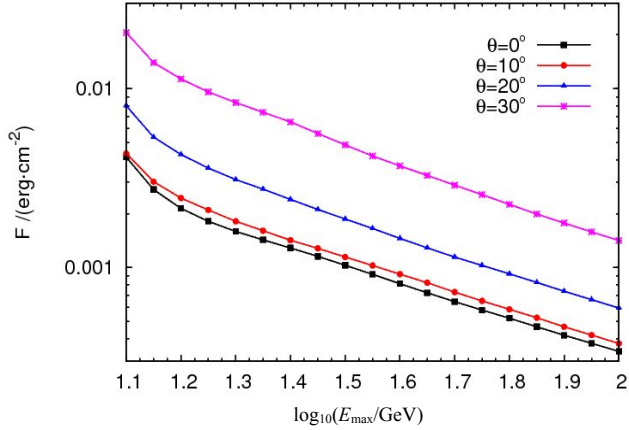

Fig. 10. The discovery fluence $F$ (for double shower front events) as a function of $E_{max}$ at $\theta = 0°$, $10°$, $20°$ and $30°$.

For comparison with double shower front events, we also simulate the sensitivity of triggered events in detecting GRBs. The effective area for triggered events is about 0.008 m$^2$ with the primary energy of 10 GeV at $\theta = 10°$, which is lower than that of double shower front events. At 100 GeV, $A_{eff}$ is about 43 m$^2$. For triggered events, the angular window with radius 2.6°[30] is used to search for a GRB, and with the time duration of 1 second, the number of background events from ARGO-YBJ data is 20.1 (with the radius of angular window chosen as 2.6°) at $\theta = 10°$ and the minimum number of signal events useful for GRB signal discovery is 36.7. Then the sensitivity of triggered events can be calculated from Eq. (4) and Eq. (5). Fig. 11 shows the discovery fluence $F$ for triggered events and double shower front events as a function of $E_{max}$ at $z = 0.0$ and $\Delta t = 1$s, where the zenith angle is chosen as 10°.

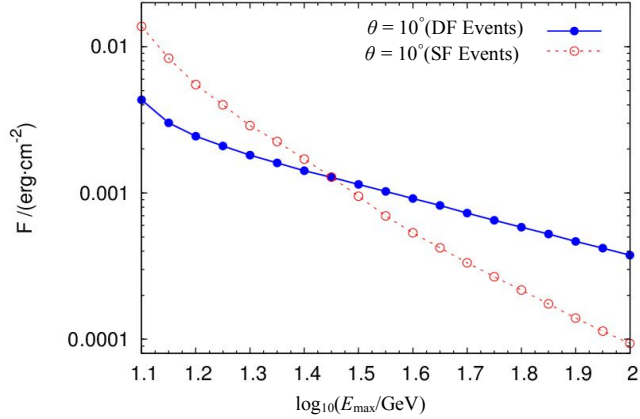

Fig. 11. The discovery fluence $F$ for both triggered events and double shower front events as a function of $E_{max}$ at $\theta = 10°$.

According to Fig. 11, the sensitivity of triggered events in detecting GRBs is worse than that of double shower front events at $E_{max} < 30$ GeV. If we analyse double shower front events and triggered events together, it will be useful to search for E > 10 GeV GRBs.

During ARGO-YBJ's lifetime, there have been some strong GRBs detected by satellite experiments, especially Fermi-LAT [6, 9]. Assuming these GRBs are in the field view of ARGO-YBJ and the power law spectra can be extended to 100 GeV, the fluence of some GRBs can be more than $10^{-4}$ erg·cm$^{-2}$. ARGO-YBJ will be sensitive enough to discover these.

## 5. Conclusions

The results from Monte Carlo simulation show that the energy threshold of ARGO-YBJ can be decreased to a few tens of GeV by using double shower front events. The angular resolution and effective area of double shower front events have been investigated by a full Monte Carlo simulation which is driven by the CORSIKA and G4argo programs. Meanwhile, the sensitivity of detecting GRBs is found to depend on the slope and energy cutoff of the spectrum, and on the redshift, time duration and zenith angle of the GRB. Then, it is further shown that for a GRB with a zenith angle around 10°, with power law index around -2.0, and energy cutoff less than 100 GeV, if the time duration lasts 1 s, the required minimum signal fluence of double shower front events is as low as $10^{-4}$ erg/cm$^2$. Under the same conditions, the sensitivity of double shower front events in detecting GRB is better than that of triggered events at $E_{max} < 30$ GeV. This work could be of great help in the analysis of ARGO-YBJ data.

*The authors would like to express their sincere thanks to the members of the ARGO-YBJ collaboration and to Prof. Ding lin-kai for many fruitful discussions and comments.*

# ARGO 实验中双前峰面事例和探测(10,100)GeV 能区γ暴灵敏度的模拟研究*

周勋秀[1; 1)]　高兰兰[1]　张宇[1]　郭义庆[2]　朱清棋[2]　贾焕玉[1]　黄代绘[1]

[1] 西南交通大学物理科学与技术学院 成都 610031

[2] 中国科学院高能物理研究所 北京 100049

**摘要：**位于西藏羊八井宇宙线观测站（海拔4300米）的ARGO-YBJ实验是全覆盖式地面宇宙线观测实验，观测γ天文的阈能~300GeV。该实验记录的大部分事例是"单前峰面"事例，需要满足阵列的触发条件（即在给定的时间窗口内至少有20个粒子击中探测器），这种事例称为"触发"事例。然而，在触发时间窗口内，其它宇宙线粒子也可能随机地击中探测器，即使这个事例不满足触发条件也将与"触发"事例同时被记录下来，这种事例称为"双前峰面"事例。由于"双前峰面"事例不需要满足触发条件，可降低探测器的阈能达几十个GeV。通过Monte Carlo模拟，本文对"双前峰面"事例的角分辨进行了详细的模拟研究，并估算了该类事例探测γ暴的灵敏度。

**关键词:** "双前峰面"事例, 角分辨, γ暴, 灵敏度, 羊八井 ARGO 实验

---